\documentclass{aa}  

\usepackage{multirow}

\usepackage{graphicx}
\usepackage{txfonts}
\begin{document}

   \title{Aperture photometry on asteroid trails:}

   \subtitle{detection of the fastest rotating near-Earth object.}

   \author{Maxime Devog\`{e}le
          \inst{1},
          Luca Buzzi
          \inst{2},
          Marco Micheli\inst{1},
          Juan Luis Cano\inst{1},
          Luca Conversi\inst{1},
          Emmanuel Jehin\inst{3},
          Marin Ferrais\inst{4},
           Francisco Oca\~{n}a\inst{1,5},
           Dora F\"{o}hring\inst{1},
           Charlie Drury\inst{1},
           Zouhair Benkhaldoun\inst{6},
           \and
           Peter Jenniskens\inst{7}
          }

   \authorrunning{Devog\`{e}le et al.}

   \institute{ESA NEO Coordination Centre, Largo Galileo Galilei, 1, 00044 Frascati (RM), Italy \\
              \email{maxime.devogele@ext.esa.int}
              \and
            Schiaparelli Astronomical Observatory, Varese, Italy
            \and
            Space sciences, Technologies \& Astrophysics Research (STAR) Institute University of Li\`{e}ge All\'{e}e du 6 Ao\^{u}t 19, 4000 Li\`{e}ge, Belgium
            \and
            Florida Space Institute, University of Central Florida, 12354 Research Parkway, Partnership 1 building, Orlando, FL 32828, USA
            \and
            ESA ESAC / PDO, Bajo del Castillo s/n, 28692 Villafranca del Castillo, Madrid, Spain
            \and
            Oukaimeden Observatory, High Energy Physics and Astrophysics Laboratory, Cadi Ayyad University, Marrakech, Morocco
            \and
            SETI Institute, 339 Bernardo Ave, Mountain View, CA 94043, USA
             }


  \abstract
   {Near-Earth objects (NEOs) on an impact course with Earth can move at high angular speed. Understanding their properties, including rotation state, is crucial for assessing impact risks and mitigation strategies. Traditional photometric methods face challenges in collecting data on fast-moving NEOs accurately.}
   {This study introduces an innovative approach to aperture photometry tailored to analyzing trailed images of fast-moving NEOs. Our primary aim is to extract rotation state information from these observations, particularly focusing on the efficacy of this technique for fast rotators.}
   {We applied our approach to analyze the trailed images of three asteroids: 2023 CX1, 2024 BX1, and 2024 EF, which were either on a collision courses or performing a close fly-by with Earth. By adjusting aperture sizes, we controlled the effective exposure times to increase the sampling rates of the photometric variations. This enabled us to detect short rotation periods that would be challenging with conventional methods.}
   {Our analysis revealed that trailed photometry significantly reduces overhead time associated with CCD read-out, enhancing the sampling rate of the photometric variations. We demonstrated that this technique is particularly effective for fast-moving objects, providing reliable photometric data when the object is at its brightest and closest to Earth.  For asteroid 2024 BX1, we detected a rotation period as short as $2.5888 \pm 0.0002$ seconds, the fastest ever recorded. Our findings underscore the efficacy of trailed observations coupled with aperture photometry for studying the rotation characteristics of small NEOs, offering crucial insights for impact risk assessment and mitigation strategies.}
   {}


   \maketitle

%

\section{Introduction} \label{sec:intro}

Circular aperture photometry is a fundamental technique, in astronomical observations to extract the flux of pointsources \citep{Howell_1989} . Its application extends across various astronomical phenomena, including the characterization of asteroid absolute magnitude and spin period \citep{Mommert_2017}. 

During the observations, asteroids are moving on the plane of the sky relative to stars and other background sources. As a consequence, to perform traditional circular aperture photometry, the exposure time has to be tuned for both the moving object and the stars to not appear elongated on the image. For most asteroids, that is not a problem. Main-belt objects rarely move at speeds larger than one arc-second per minute ($''$/min). 

In the case of near-Earth objects (NEOs; asteroids whose perihelion distance is smaller than $q=1.3$ astronomical units), traditional aperture photometry encounters unique challenges due to the rapid motion of these objects against the sky. Indeed, their motion is usually on the order of several $''$/min and can reach very high speeds when performing a close Earth fly-by. Such high angular speed requires short exposure times, leading to most of the observing time being spent on CCD read-out rather than on-sky observations. This has prevented the detection of very fast rotation periods. The fastest rotation period, of P = 2.99 s, was detected using a fast read-out camera \citep{Beniyama_2022}.

\begin{figure}
\centering
\includegraphics[width=9.24cm]{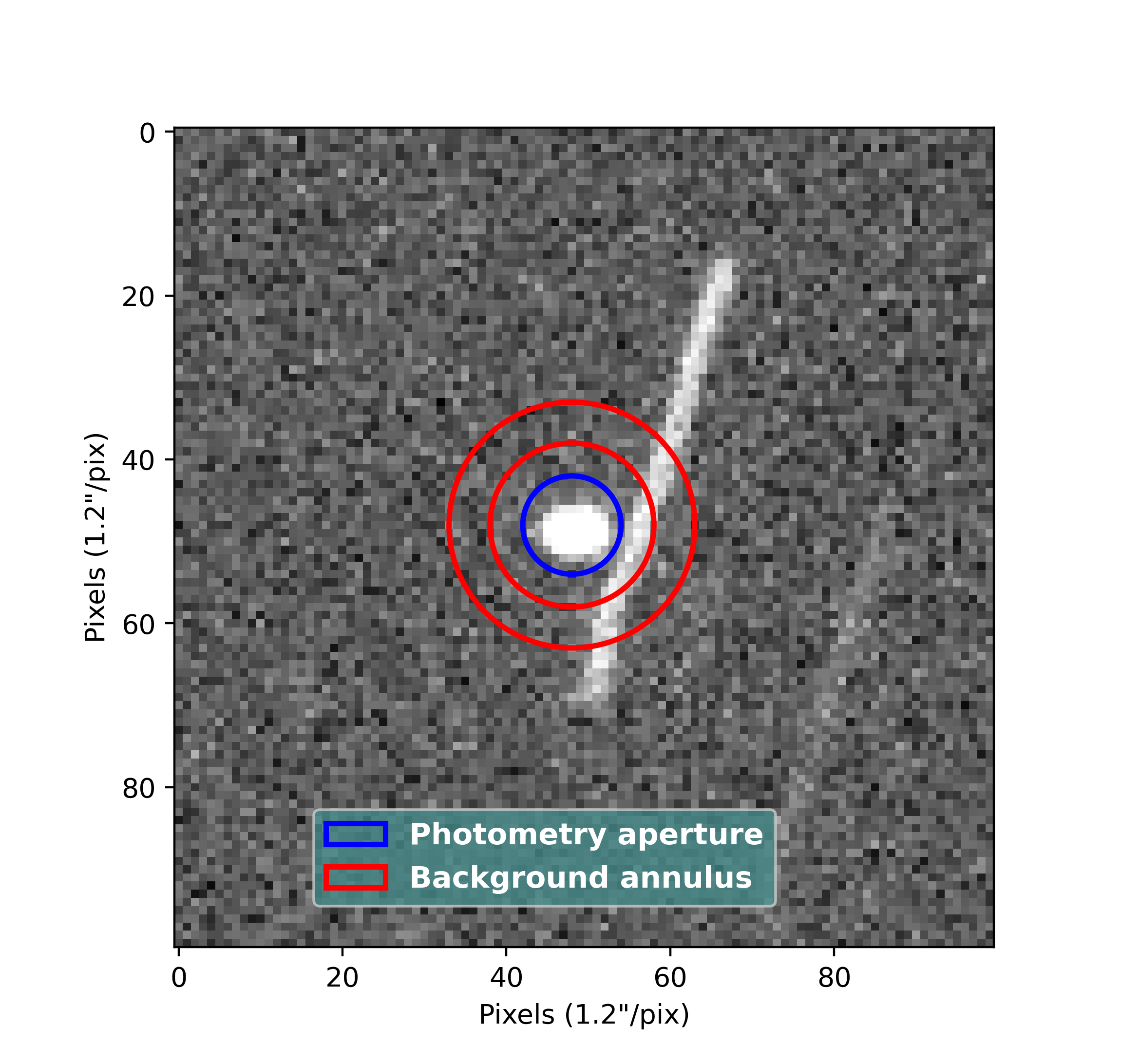}
\caption{Example of circular aperture photometry on a 10-second observation of 2024 EF obtained with the TRAPPIST-North telescope (pixel size of $1.2''$/pix). The flux of the target is captured by the first innermost blue aperture, while the gap between the innermost and second aperture is ignored. The pixels between the outermost and second aperture (two red apertures) are used to compute the sky-background level. }
\label{fig:Circ_Aper}
\end{figure}

This predicament becomes especially pronounced for newly discovered objects that are approaching the Earth on a close fly-by or collision trajectory. Poorly known ephermerides can makes these objects challenging to point and track. Not all observatories are equipped to point and follow undesignated objects. Newly discovered objects initially appear on the NEO Confirmation Page\footnote{\url{https://www.minorplanetcenter.net/iau/NEO/toconfirm_tabular.html}}before being published in a Minor Planet Electronic Circular and receiving a formal designation. Regular minor-body ephemeris services, on which most observatories rely, do not provide ephemerides for these undesignated objects. Their rapid motion and variation in motion over short periods of time, which are often shorter than the typical exposure time, present challenges for observatories. Some observatories need to manually provide motion rates and cannot adjust them during the observations. The response time and observation window for these objects are usually on the order of hours. This does not leave the observer enough time to experiment with the observing strategy. For these reasons, and because fast read-out cameras are rarely available, even traditional tracked observations with short exposure times can lead to both the stars and asteroid appearing trailed in the image. These observations are hard to analyse and even if the asteroid is perfectly tracked, the trailed star images can overlay the asteroid image or the selected background field, making the measurement useless (see Fig. \ref{fig:Circ_Aper}). In this example, the observation is performed tracking the asteroid. Here, imperfections in the ephemerides computation cause the asteroid to appear trailed differently to the stars, resulting in sub-optimal measurements of both the stars and the asteroid flux. Moreover, their close distance to Earth makes even small asteroids unusually bright allowing to obtain high signal-to-noise ratio on the asteroid trailed images.

Therefore, the easiest and most reliable technique for obtaining high-quality data on these objects, when only slow read-out time CCD is available, is to perform sidereally tracked observations, allowing the asteroid to trail in the images. 

In previous work, photometry and astrometry on trailed observations of near-Earth asteroids was mostly focused on fitting the whole trail \citep{Veres_2012}. More recently, \citet{Bolin_2024} intentionally trailed their object and controlled the drift rate. This makes the stars to also appear trailed on the image. 

\begin{figure}
\centering
\includegraphics[width=8.76cm]{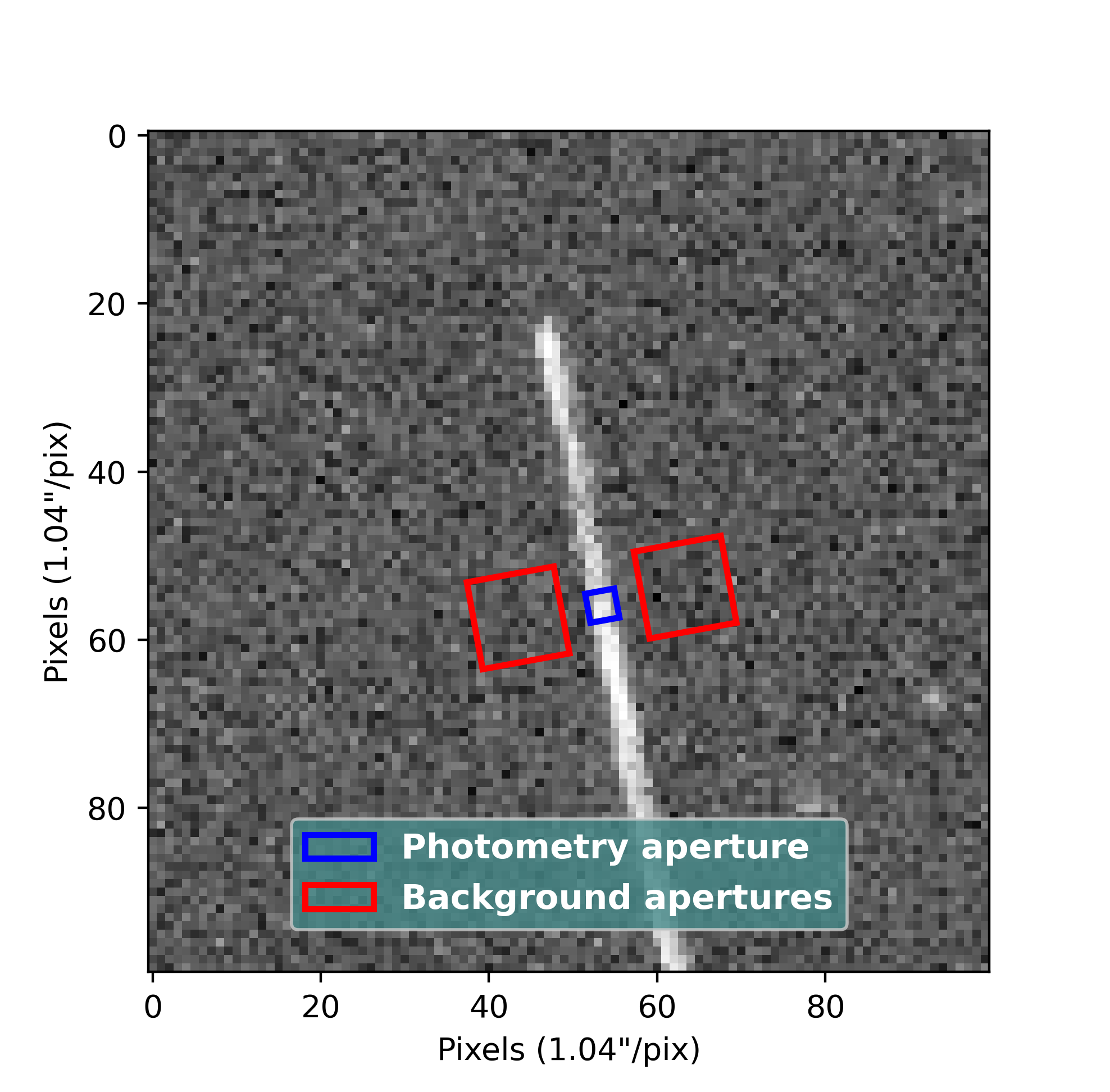}
\caption{Example of our square apertures on a trailed observation of 2024 BX1 (pixel scale of $1.04''$/pix) observed at the Schiaparelli Observatory. The blue square aperture centered on the trail is used to capture the asteroid flux. The red square apertures located on both sides of the trail are used to compute the sky-background level.}
\label{fig:Aperture_Square}
\end{figure}

Here, we present an aperture photometry technique that does not require a precise fit of the trail. In our approach, we take advantage of the fast speed of the object on the sky and keep the stars sharp. This allows, in our case, the use of regular circular apertures for the stars, and we can therefore use them to calibrate the field both photometrically and astrometrically. The method allows us to measure the photometric properties from images where the object's trajectory extends across hundreds of pixels. 

The method is applied to three recently observed NEOs with trailed PSFs. Of these, 2023 CX1 and 2024 BX1 impacted the Earth on 2023 February 13 and 2024 January 21 respectively. The third one, 2024 EF, performed a close fly-by at a distance of only $57\,614.5 \pm 2.4$ km from the Earth's center on 2024 March 4. All of these asteroids are small, with $H$ magnitudes ranging from $H=29.1$ for 2024 EF to $H=32.7$ for both 2023 CX1 and 2024 BX1. These magnitudes correspond to sizes ranging from less than 0.5 m \citep{Spurny_2024} to approximately 5 m, enabling them to display very fast rotation periods \citep{Beniyama_2022,Thirouin_2016,Thirouin_2018}. 

\section{Aperture photometry on trailed objects}

Aperture photometry on trailed observations of fast moving asteroids represents a significant departure from conventional techniques, such as aperture photometry on point-source like objects. In traditional aperture photometry, circular apertures are typically employed to enclose the target positions in the image and define the background sky brightness, facilitating the measurement of its flux. 

To perform aperture photometry on trailed observation of asteroids, we use a square or rectangular aperture aligned with the direction of the NEO's trail (see Fig. \ref{fig:Aperture_Square}). This technique maximizes the signal-to-noise ratio of the extracted photometry over a small section of the trail. We then step along the trail to collect the photometry as a function of time.

As conventionally used for circular aperture photometry, we estimate the flux of the sky-background at an appropriate distance from the observed trailed object allowing to accurately assess the background flux affecting the source without being contaminated by it. We use the \textit{Python} package \textit{Photutils} \citep{Photutils} to generate the rectangular apertures and conducting the photometry. We also utilize the source masking capabilities of \textit{Photutils} to remove pixels contaminated by background sources within the aperture used to estimate the sky-background flux.  

To extract temporal information from the observed trails, we first perform astrometric calibration using the stars. This step is performed using the \textit{photometrypipeline} \textit{Python} package \citep{Mommert_2017} that is automatically extracting sources using \textit{sextractor} \citep{Bertin_1996} and then solving the plate astrometrically using \textit{scamp} \citep{Bertin_2006}. The sidereally tracked observations allow to obtain accurate plate solution. Next, we fit both ends of the trail to precisely locate it on the image and calculate its length. We divide the temporal span between the beginning and the end of the trail into uniform time intervals, adjusting the aperture dimensions to account for the object's changing speed during acquisition. The final step involves multiplying the observed flux for each individual aperture by the ratio between the total and effective exposure time (time needed for the target to cross the aperture). This is done to obtain calibrated magnitudes for the asteroid observations. 

The effective exposure time for each aperture can be chosen by the user. However, the optimum value is typically chosen to correspond to aperture sizes approximately equal to the Full Width at Half Maximum (FWHM) of the circular stars \citep{Mighell_1999}. For each time interval the location of the object on the image is computed using ephemerides. The final position for each aperture is then determined by fitting one dimensional Moffat function \citep{Moffat_1969} to the trail.

The optimum apertures are square apertures whose sizes are approximately equal to the FWHM computed using the stars. This ensures both the optimal extraction of the flux perpendicular to the trail \citep{Mighell_1999}, but it also provides the smallest apertures that could be used to obtain independent temporal measurements. Here, for two consecutive measurements, we use apertures that touch each other, but without overlaps, to maximize the flux extraction. This makes consecutive observations slightly correlated with each other. When the target is centered at the limit of two square apertures, half of its PSF will be located in one aperture and the other half will be located in the other aperture. To obtain fully independent observations, gaps of at least one FWHM need to be added between measurements. Alternatively, longer rectangular apertures can be used in the direction of the trail to increase the flux, thus increasing the signal-to-noise ratio of individual measurements at the cost of decreasing the temporal resolution. 

It is important to note that we rely on the sky to be photometric during the exposure time as we cannot use stars to calibrate the temporal variations observed along the trail. However, the same issue arises for star-trailed observations.  

\begin{figure}
\centering
\includegraphics[width=9cm]{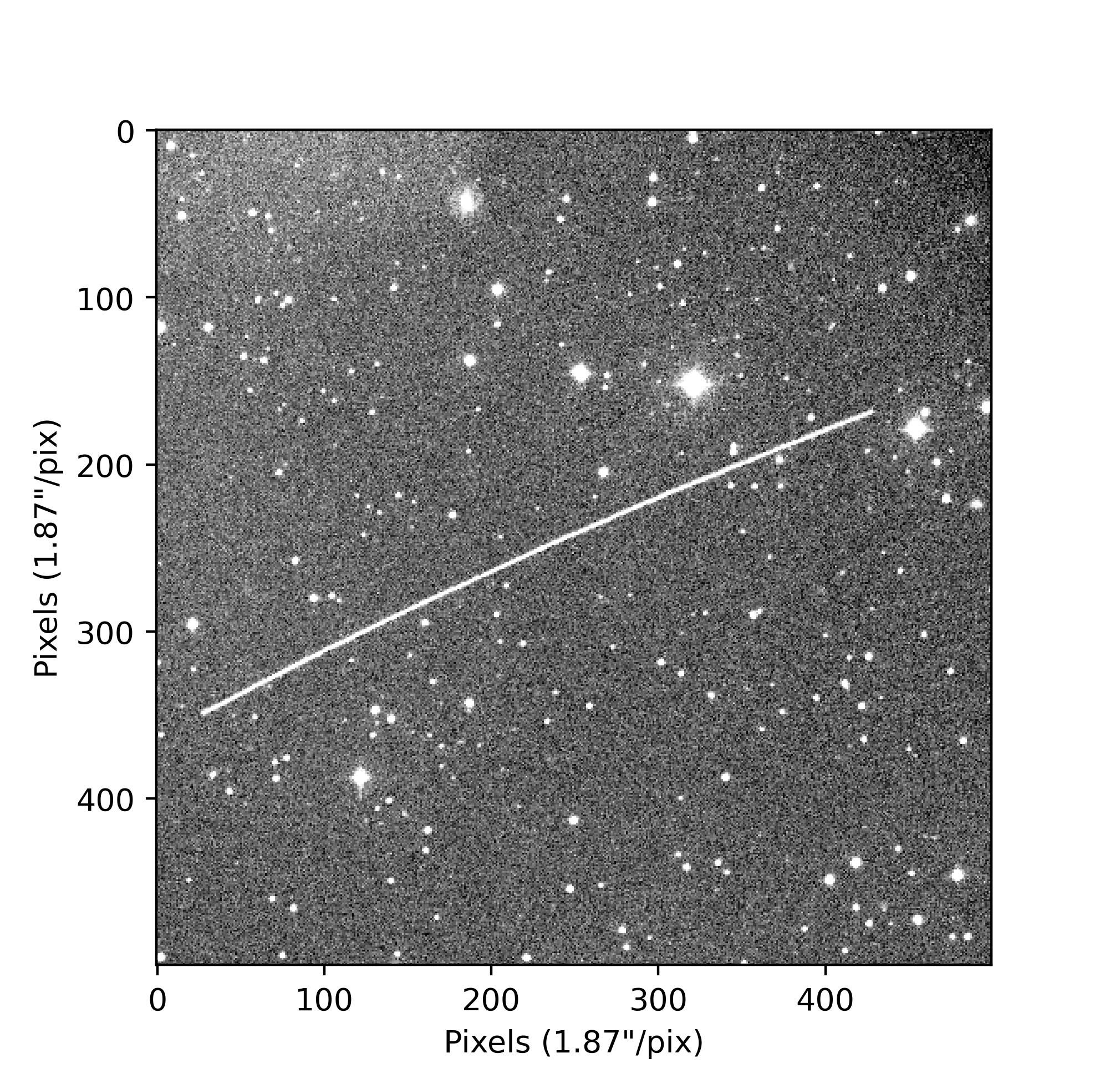}
\caption{Observation of 2023 CX1 nine minutes before impact (pixel scale of $1.87''$/pix) obtained at the Schiaparelli Observatory. The exposure time was 60 seconds, during which CX1 can be seen trailing over 441 pixels. The curvature of CX1's trail is attributed to its very close proximity to the observer (approximately 7\,000 km) and the rapid variation of its relative motion in the sky.}
\label{fig:trail_2023CX1}
\end{figure}

\begin{figure}
\centering
\includegraphics[width=9cm]{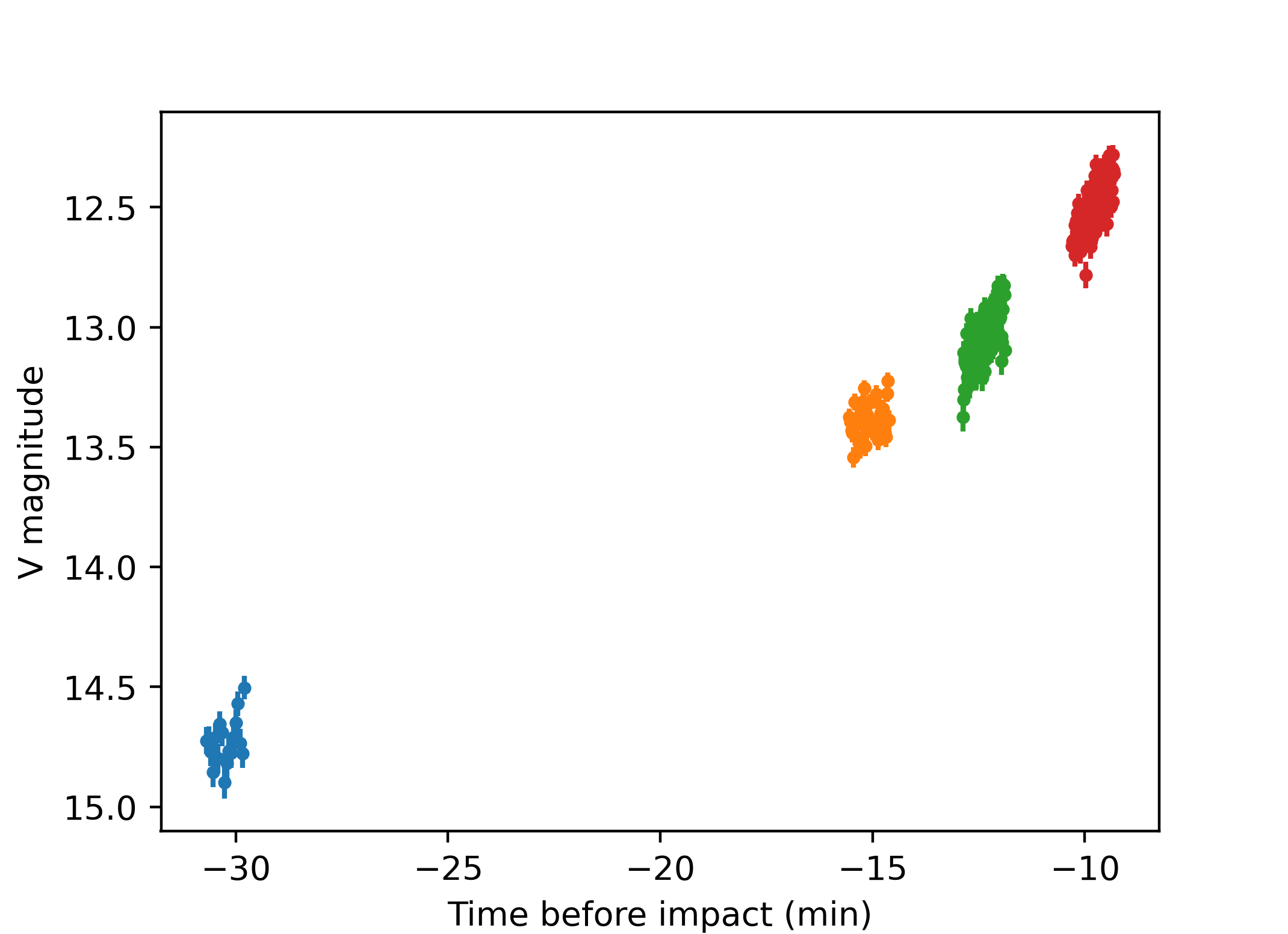}
\caption{2023 CX1 photometry of the four trailed images. Each color represents a different image of 60 s each. The $x$ axis represents the time in minutes before the impact time while the $y$ axis is the V magnitude.}
\label{fig:Photometry_CX1}
\end{figure}

\section{Observations and results}
\subsection{2023 CX1}

The asteroid 2023 CX1 (hereafter CX1), also temporarily known with the discoverer-assigned identificator Sar2667, was initially spotted by Krisztián Sárneczky, an astronomer at the Konkoly Observatory's Piszkéstető Station (MPC code 461) in Hungary. This observation was obtained on 2023 February 12 at 20:18 UT and quickly triggered impacting alerts from the NASA and ESA impacting monitoring services Scout\footnote{\url{https://cneos.jpl.nasa.gov/scout/intro.html}} and Meerkat \citep{Fruhauf_2021}. Six hours after its discovery, CX1 impacted on 2023 February 13 at 02:59 UT, over the coast of Normandy in France. This event marked the seventh time an asteroid had been discovered before impacting Earth. By combining the precise knowledge of its trajectory while entering the atmosphere with atmospheric wind data and the observed fragmentation altitude from fireball observations in the FRIPON meteor monitoring network \citep{Colas_2020}, meteorites were recovered on the ground \citep{Jenniskens_Colas_2023}. This was only the third time that meteorites were recovered from an asteroid discovered prior to impacting Earth \citep{Jenniskens_2009, Jenniskens_2021}. 

\begin{figure*}
\centering
\includegraphics[width=18cm]{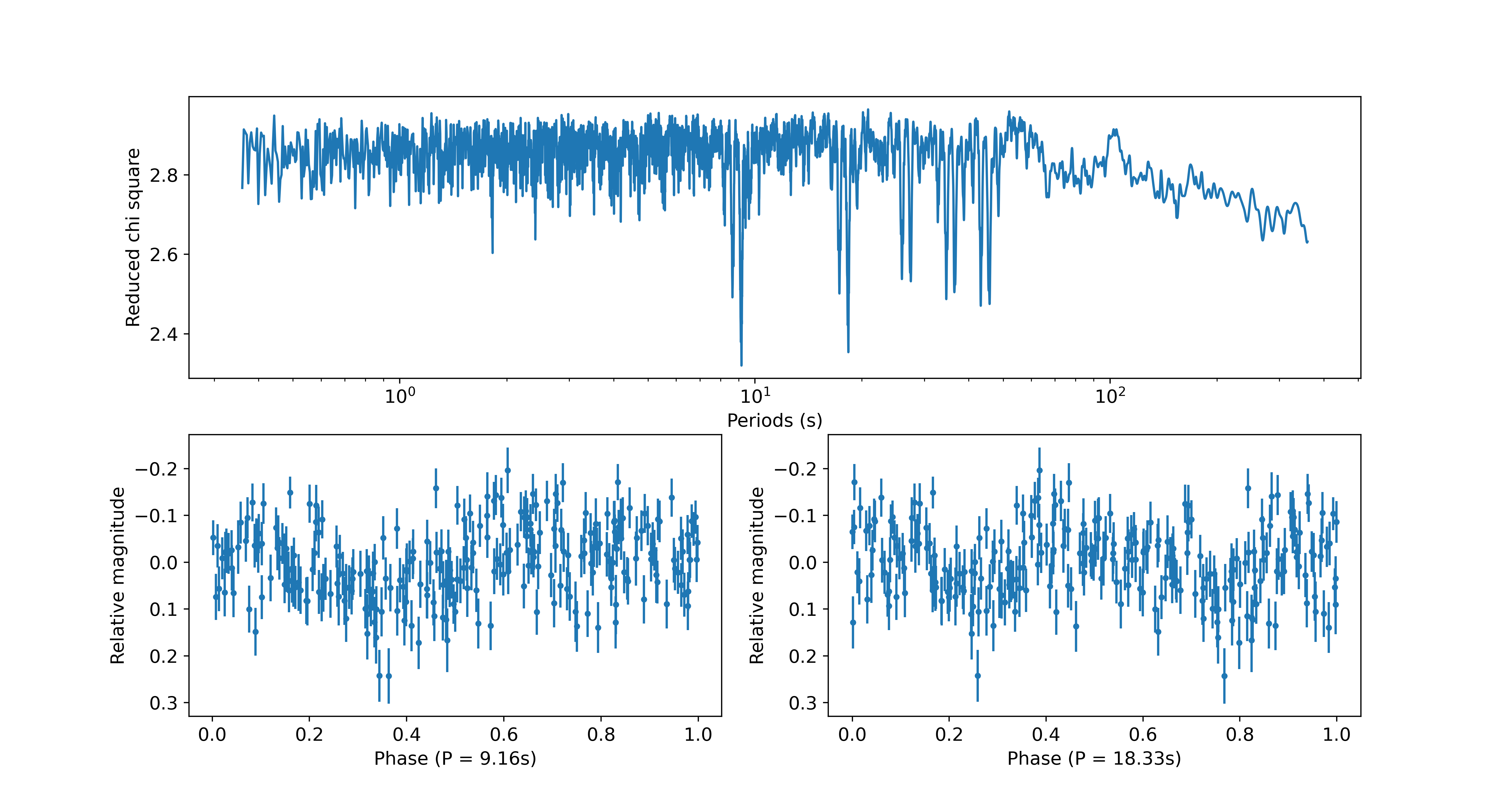}
\caption{Upper plot: Periodogram for all the observations of 2023 CX1, testing periods between 0.36 s to 6 min. Bottom plots: Observations phased according to the two best test periods. We do not observe any convincing solution for the rotation period of 2023 CX1 although the solutions at P=9.16 s (left) and 18.33 s (right) are plausible.}
\label{fig:Phase_CX1}
\end{figure*}

At the time of its discovery, the object was already within $233\,000$ km of Earth and moving at a speed exceeding $11''/$min. Over the next six hours its speed gradually increased until more than $10\,000''$/min, just before entering the Earth shadow and impacting. With an estimated absolute $H$ magnitude of $32.7$, its size is estimated to range between 0.8 to 1.7 m. 

We analyse four observations obtained at the Schiaparelli Observatory located in northern Italy, atop mount Campo dei Fiori near Varese, 1230~m above sea level (MPC 204). A 0.84~m Newtonian telescope was used with an SBIG STX-16803, which provided a field of view of $42'\times42'$ with a pixel size of $1.87''$ when operated in $3\times3$ binning mode. The observations were unfiltered, with an exposure time of 60 s. 

The first observation was obtained at 02:29 UT when CX1 was moving at a speed of $\sim147 ''$/min (varying significantly over the time span of the exposure) and thus trailing over 76 pixels on the images. The last image was obtained at 02:50 UT, only 9 min before impact, when CX1 was located at 7\,000 km from the observer (varying between 7\,575 km and 6\,877 km during the 60 s of the exposure). At that time, it was moving at a speed of more than $900''$/min trailing over 441 pixels. The rapid change in Earth's ranging and the relative motion in the sky cause the trail to curve significantly as can be seen in Fig. \ref{fig:trail_2023CX1}. Information on all the individual exposures can be found in Table \ref{tab:Observations}.

The CX1 photometric observations are presented in Fig. \ref{fig:Photometry_CX1}, with each color representing a different observation. The magnitude has been calibrated in the $V$ band independently for each acquisition using the field stars. The time is expressed in minutes before the impact. The fast brightening of CX1 is clearly visible as it approaches Earth.  

We searched for the signature of rotation by folding all the data according to trial periods. For each period, a Fourier series of order 5 is fitted and the chi-square of the fit to the data is computed. The phase curve is expected to have two minima for one rotation, but can be complicated if the object is tumbling. Fig. \ref{fig:Phase_CX1} shows the periodogram of CX1 for periods between 0.36 seconds and 6 minutes. The periodogram exhibits several chi-square minima. 

Fig. 5 shows two attempts to fold the data according to the two strongest minima. A period of 9.16 s (left diagram) does not show the expected double minimum in the phased lightcurve, which may represent just half a period. A period of 18.33 s does result in the expected double minimum, but the amplitude is small leading to poor significance and likely due to noise rather than the asteroid rotation.

The photometric accuracy is approximately 0.1 mag, with an effective exposure time around 0.25 s. In this case, the asteroid is nearly spherical in shape, was observed close to a pole-on geometry, or has a period much longer than 6 min, resulting in a lightcurve amplitude smaller than 0.1 mag. 

\begin{figure*}
\centering
\includegraphics[width=18cm]{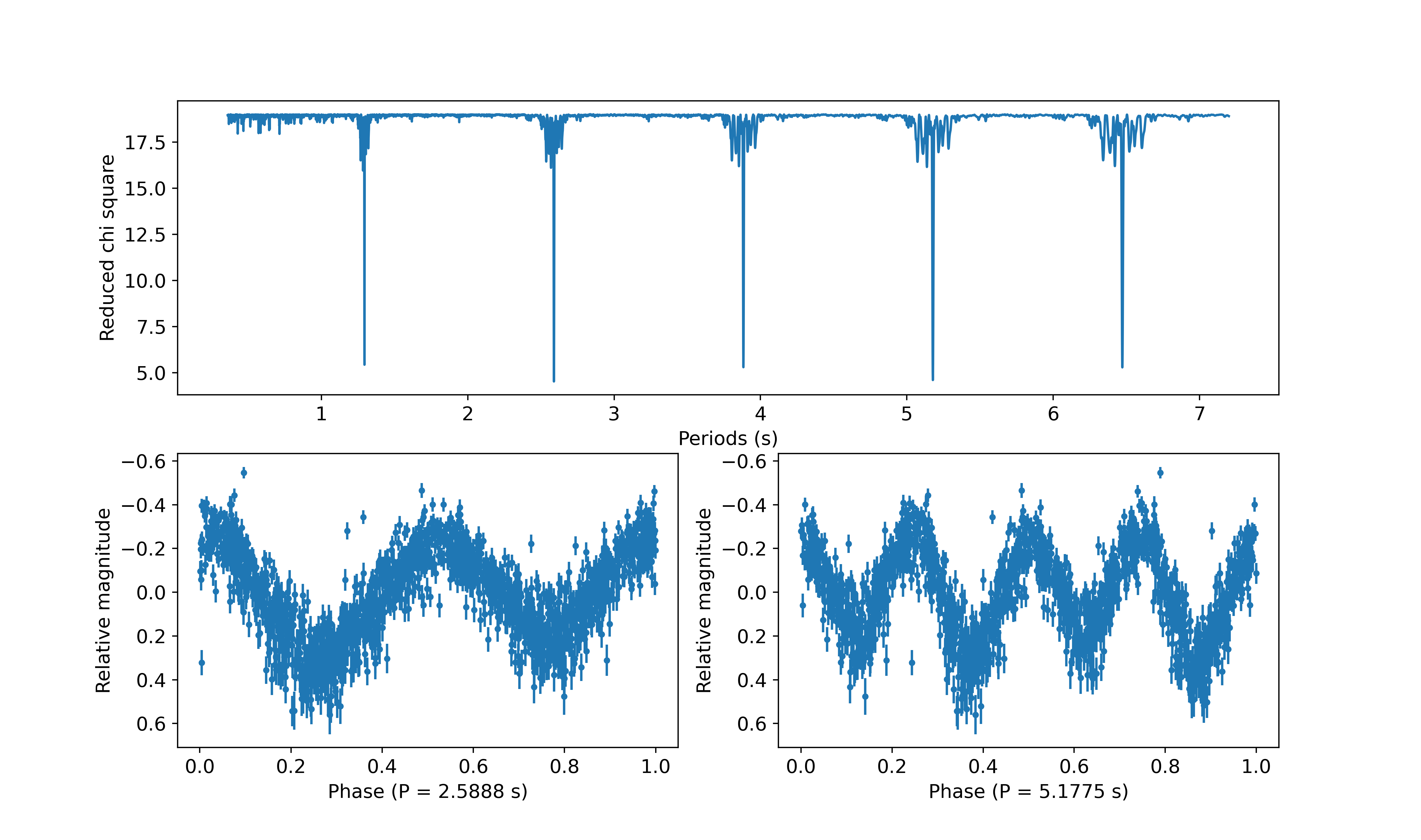}
\caption{Upper plot: Periodogram of all the observations of 2024 BX1 testing periods between 0.36 to 7.2 s. We see a clear minimum at 2.5888 s and its aliases. Bottom: Observations phased according to the two best results, 2.5888 and 5.1775 s respectively.}
\label{fig:2024BX1_Phase}
\end{figure*}

\begin{figure}
\centering
\includegraphics[width=9cm]{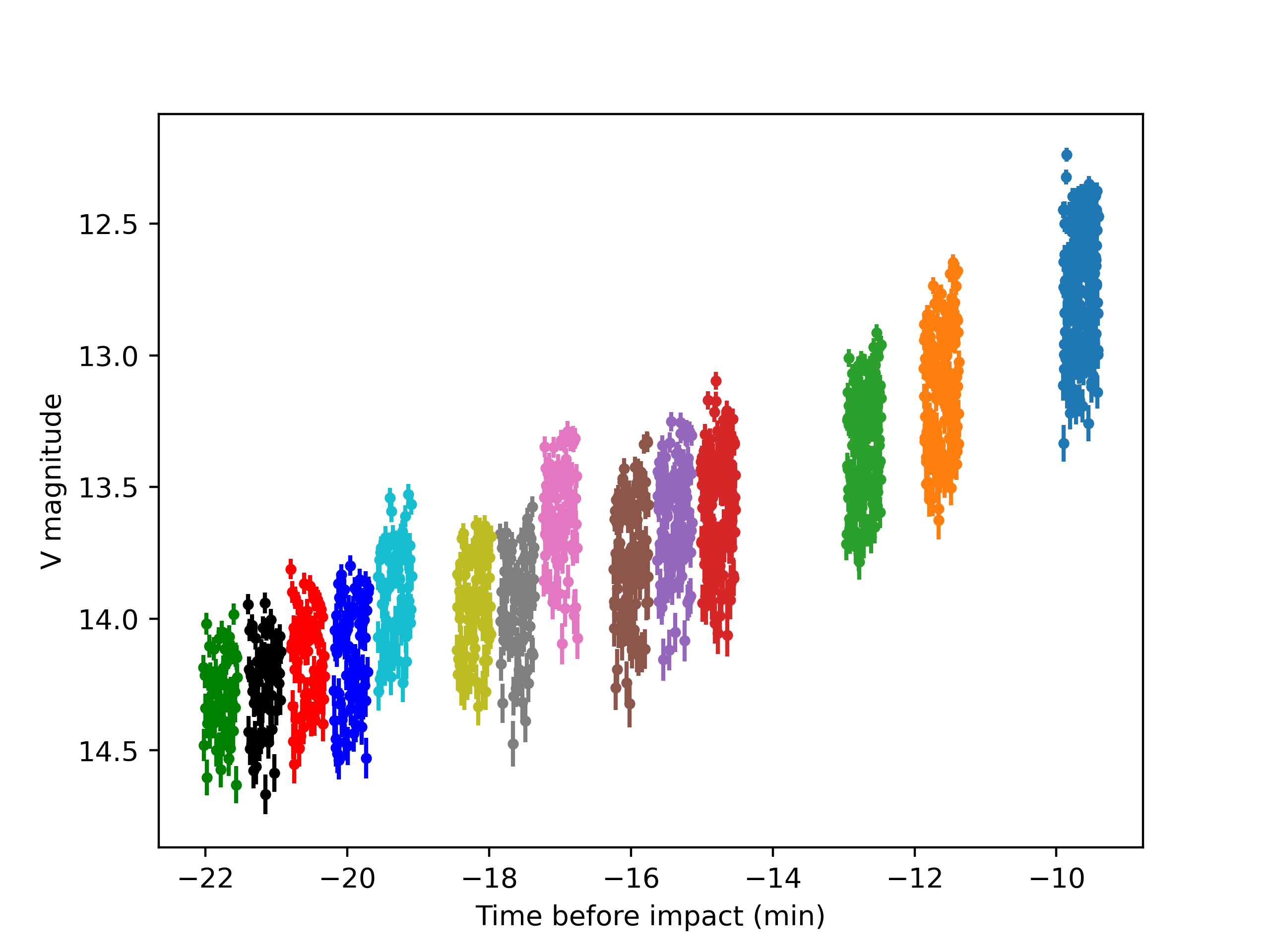}
\caption{Photometry of 2024 BX1 on all the trailed images. Each color represents a different image of 30 s each. The $x$-axis represents the time in minutes before the impact time while the $y$-axis is magnitude in the $V$ band.}
\label{fig:2024BX1_All}
\end{figure}

\subsection{2024 BX1}

2024 BX1 (hereafter BX1), temporarily labeled as Sar2736 by the discoverer, is the third impactor asteroid discovered by Krisztián Sárneczky at the Hungary's Konkoly Observatory's Piszkéstető Station (MPC code 461). As with 2023 CX1, the discovery occurred just a few hours prior to its impact on Earth, leaving very little time for its physical characterization. Meteorites were recovered on the ground \citep{Jenniskens_2024}. We analyse here trailed observations obtained at the Schiaparelli Observatory. Like CX1, BX1 is displaying the same $H$ magnitude of 32.7, leading to the same estimated size of 0.8 to 1.7 m. Analysis of the fireball lightcurve suggests a smaller size of 0.44 m, probably because the aubrite meteorites recovered on the ground display an unusually high albedo \citep{Jenniskens_2024a, Spurny_2024}.

The telescope used is the same as for CX1, but with a QHY600M CMOS camera instead of the SBIG STX-16803. The QHY600M offers a $41.6'\times27.8'$ field of view with a pixel scale of $1.04''$ when used in a 4 by 4 binning mode. The first observations were obtained when the asteroid was located at a distance of 17\,500 km from Earth and moving at a speed of $7.6''$/s, while the last observations were obtained just 11 minutes before impact at a distance of 9\,500 km from the observer and moving at a speed of more than $30''$/s. Information about all the individual observations can be found in Table \ref{tab:Observations}

Contrary to CX1, we find a clear rotational signature at a period of $P = 2.5888 \pm 0.0002$ s, displaying a combined phase curve with an amplitude of 0.7 mag (Fig. \ref{fig:2024BX1_Phase}).  All the photometric observations, calibrated in the $V$ band independently for each acquisition using the field stars, are combined in Fig. \ref{fig:2024BX1_All}, with the $x$ axis representing the time before impact. As for CX1, we again see a clear brightening of the object as it approached Earth. Fig. 6 shows the periodogram (the same method as for 2023 CX1 is used to compute the periodogram) for test periods between 0.36 to 7.2 s. The signal at a period of $P = 2.5888 \pm 0.0002$ s, along with its aliases (varying numbers of maxima and minima), is evident.

This is the fastest rotation period ever measured for an asteroid, $13\%$ faster than the previous record held by 2020 HS7 with a rotation period of  $P = 2.99$ s \citet{Beniyama_2022}. Analysis of the signification of such a fast rotation is outside the scope of this paper and is left to other work. 


\begin{figure*}
\centering
\includegraphics[width=18cm]{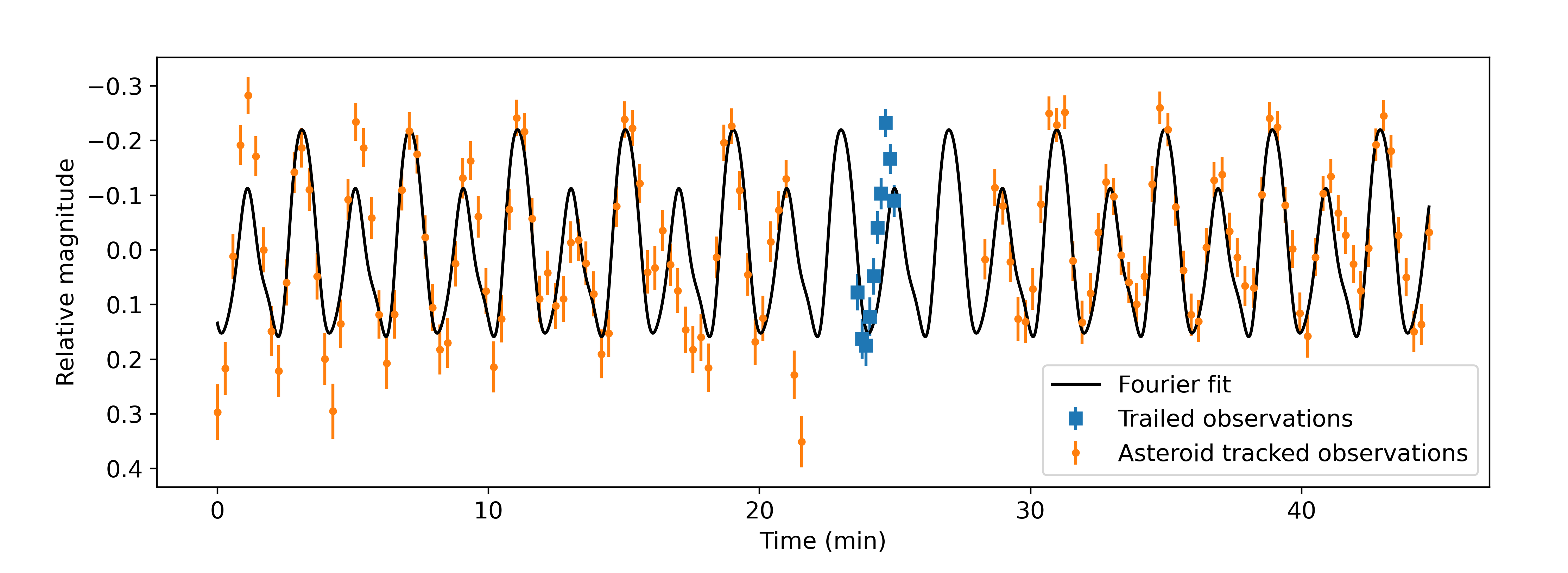}
\caption{Ligthcurve of 2024 EF from 02:43 UT to 03:33 UT on 2024 February 4 obtained with TRAPPIST-North. The orange dots corresponds to regular asteroid tracked observations, the blue squares corresponds to the photometry obtained in one single trailed observations of 90 s. The black curve corresponds to the best Fourier fit on the orange observations only.}
\label{fig:2024EF_Both}
\end{figure*}

\begin{figure}
\centering
\includegraphics[width=9cm]{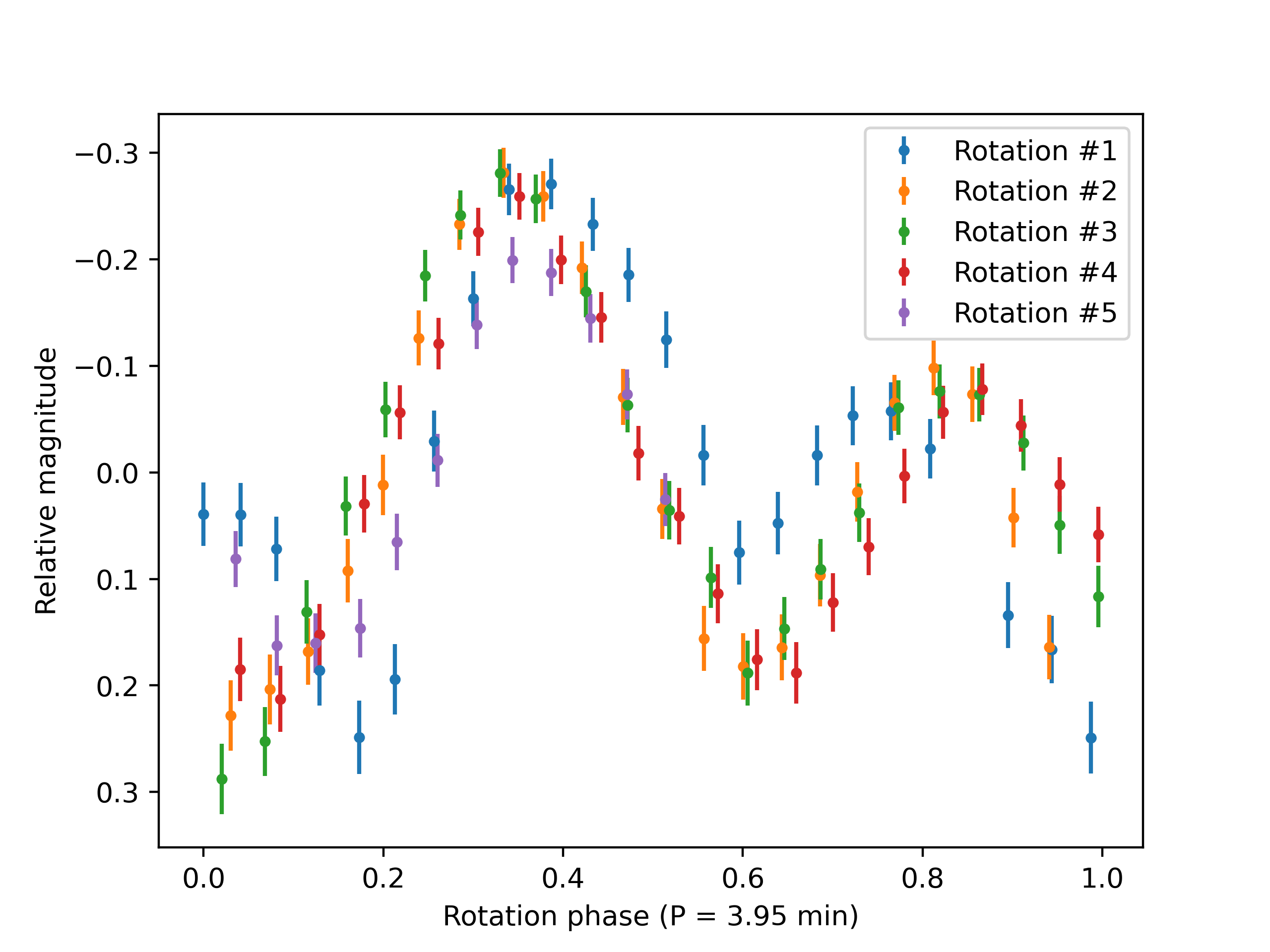}
\caption{Phased lightcurve of 2024 EF according to a period of 3.95 min.}
\label{fig:2024EF_Period}
\end{figure}

\subsection{2024 EF}

The trailed imaging technique is not limited to small Earth impactors when they are close to Earth. Significantly larger asteroid 2024 EF (hereafter EF) performed a close fly-by with Earth at a minimal distance of $57\,614.5 \pm 2.4$ km on 2024 March 4, less than 48 hours after its discovery on March 2. With an estimated $H$ magnitude of 29, its size ranges between 4 to 10 m. 

We observed EF with the TRAPPIST-North telescope located at the Oukaimaiden observatory in Morocco (TN; MPC Z53). TRAPPIST-North is a robotic telescope using a 0.6~m Ritchey-Chr\'{e}tien designs operating at f/8 on a German Equatorial mount. It is a clone of TRAPPIST-South located in la Silla observatory in Chile \citep{Jehin_2011}. The camera is an Andor IKONL BEX2 DD ($0.60''$/pixel, $20'\times20'$ field of view). Images were obtained with a binning of 2$\times$2 and a blue cutting filter with transmission starting at 500 nm. 

We performed both ``regular'' (observations with tracking on the asteroid motion) and trailed observations. The regular observations were of 20, 10, and 3 s and the stars were trailing over several dozens of pixels preventing from using them for photometric calibration (see Fig. \ref{fig:Circ_Aper}). However, typical circular aperture photometry allowed us to determine a rotation period of 3.95 min for this object using the latest and shortest exposure time (3 s) observations. 

We also analyze one trailed acquisition lasting 90 s. Although this acquisition is much shorter than the known rotation period, we were able to observe a trend in the photometry that could be phased with the lightcurves obtained just before and after. Fig. \ref{fig:2024EF_Both} shows the trailed observation analysis in blue squares. The photometry obtained just before and after using regular asteroid tracked observations are shown in orange. We utilized the orange observations to search for the best Fourier series fit, represented by the black continuous line. We can see that the trailed observation fits nicely the trend obtained with the orange observations only. The trailed observations have been binned 15 times to better align with the exposure times of the regular observations and to enhance visibility on the plot. 

Additionally, we note that there is a hint that EF is in a tumbling state as we can see that the amplitude of the lightcurve appears to change over time following a regular decrease and increase. 
We were also not able to phase all the observations perfectly (spanning several hours) using the 3.95 min because of the tumbling nature of the object. However, short section can be phased without issues, but we are again noticing departure from a perfect phasing as it can be seen in Fig. \ref{fig:2024EF_Period}, where each individual rotation is displayed using a different color.  

We are not reporting the uncertainty on its period as this would require a more detailed analysis of the tumbling nature of the rotation state. That work is outside of the scope of this paper.

\begin{table*}
\caption{Summary table of all the trailed observations analyzed in this work.}
    \centering
    \begin{tabular}{|c|c|c|c|c|c|c|c|c|}
    \hline
         Object                   & Observatory &  Frame               &length & Time     & RA         & DEC         & Speed & Range\\
                                  & &                       & pixels & hh:mm:ss & hh:mm:ss   & dd:mm:ss & $''$/s & km\\
        \hline
        \multirow{8}{*}{2023 CX1} & \multirow{8}{*}{Schiaparelli Observatory} &    \multirow{2}{*}{1} & \multirow{2}{*}{77} & 02:29:40 & 07:58:08.4 & +37:45:40.9  & 2.3 & 20\,819\\
        \cline{5-9}
                                 & &                        &                     &  02:30:40 & 07:58:11.5 & +37:47:52.2  & 2.4 & 20\,228\\
        \cline{3-9}

                                 & & \multirow{2}{*}{2}     & \multirow{2}{*}{168} & 02:44:49 & 07:58:09.7 & +38:34:24.0 & 5.0 & 11\,174\\
        \cline{5-9}
                                 & &                        &                      & 02:45:49 & 07:57:57.9 & +38:38:59.7 & 5.5 & 10\,517 \\
        \cline{3-9}

                                 & & \multirow{2}{*}{3}     & \multirow{2}{*}{246}  & 02:47:32 & 07:57:25.4 & +38:47:40.2 & 7.0 & 9\,345 \\
        \cline{5-9}
                                 & &                        &                       & 02:48:32 & 07:56:57.5 & +38:52:52.7 & 8.4 & 8\,673 \\
        \cline{3-9}
                                 & & \multirow{2}{*}{4}     & \multirow{2}{*}{441}   & 02:50:06 & 07:55:50.2 & +39:01:36.8 & 11.9 & 7\,578 \\
        \cline{5-9}
                                 & &                        &                        & 02:51:06 & 07:54:46.3 & +39:07:14.9 & 15.6 & 6\,874 \\
        \hline
        \multirow{28}{*}{2024 BX1} & \multirow{28}{*}{Schiaparelli Observatory} & \multirow{2}{*}{1}     & \multirow{2}{*}{220}   & 00:09:58     & 07:52:37.5       & +48:20:53.8          & 7.5     &  17\,550   \\
        \cline{5-9}
                                 & &                        &                      &    00:10:28       &  07:52:42.2       &  +48:24:29.2        & 7.7     &  17\,205   \\
        \cline{3-9}
                                 & & \multirow{2}{*}{2}     & \multirow{2}{*}{230}   &    00:10:35       &    07:52:43.6       &   +48:25:33.1        & 7.8     &  17\,105   \\
        \cline{5-9}               
                                 & &                        &                       &    00:11:05      &     07:52:48.7      &     +48:29:21.6       &  8.2    &  16\,756   \\
        \cline{3-9}
                                 & & \multirow{2}{*}{3}     & \multirow{2}{*}{243}   &    00:11:11       &    07:52:49.9       &   +48:30:19.8        & 8.2     &  16\,669   \\
        \cline{5-9}               
                                 & &                        &                       &    00:11:41      &     07:52:55.3      &     +48:34:21.6       &  8.6    &  16\,318   \\
        \cline{3-9}
                                 & & \multirow{2}{*}{4}     & \multirow{2}{*}{257}   &    00:11:48       &    07:52:56.6       &   +48:35:23.8        & 8.7     &  16\,230   \\
        \cline{5-9}               
                                 & &                        &                       &    00:12:18      &     07:53:02.2      &     +48:39:37.3       &  9.1    &  15\,880   \\
        \cline{3-9}
                                 & & \multirow{2}{*}{5}     & \multirow{2}{*}{269}   &    00:12:25       &    07:53:03.7       &   +48:40:47.1        & 9.2     &  15\,786   \\
        \cline{5-9}               
                                 & &                        &                       &    00:12:55      &     07:53:09.4      &     +48:45:11.3       &  9.6    &  15\,441   \\
        \cline{3-9}
                                 & & \multirow{2}{*}{6}     & \multirow{2}{*}{301}   &    00:13:33       &    07:53:17.5       &   +48:51:21.5        & 10.2     &  14\,980   \\
        \cline{5-9}               
                                 & &                        &                       &    00:14:03      &     07:53:23.9      &     +48:56:16.5       &  10.7    &  14\,631   \\
        \cline{3-9}
                                 & & \multirow{2}{*}{7}     & \multirow{2}{*}{318}   &    00:14:09       &    07:53:25.5       &   +48:57:34.7        & 10.8     &  14\,541   \\
        \cline{5-9}               
                                 & &                        &                       &    00:14:39      &     07:53:32.3      &     +49:02:50.0       &  11.3    &  14\,188   \\
        \cline{3-9}
                                 & & \multirow{2}{*}{8}     & \multirow{2}{*}{341}   &    00:14:46       &    07:53:34.1       &   +49:04:14.3        & 11.5     &  14\,097   \\
        \cline{5-9}               
                                 & &                        &                       &    00:15:16      &     07:53:41.4      &     +49:09:52.7       &  12.1     &  13\,741   \\
        \cline{3-9}
                                 & & \multirow{2}{*}{9}     & \multirow{2}{*}{379}   &    00:15:45       &    07:53:49.2       &   +49:15:59.8        & 12.7     &  13\,374   \\
        \cline{5-9}               
                                 & &                        &                       &    00:16:15      &     07:53:57.2      &     +49:22:16.5       &  13.5    &  13\,016   \\
        \cline{3-9}
                                 & & \multirow{2}{*}{10}     & \multirow{2}{*}{409}   &    00:16:22       &    07:53:59.3       &   +49:23:54.2        & 13.6     &  12\,926   \\
        \cline{5-9}              
                                 & &                        &                       &    00:16:52      &     07:54:07.9      &     +49:30:41.5       &  14.4    &  12\,563   \\
        \cline{3-9}
                                 & & \multirow{2}{*}{11}     & \multirow{2}{*}{436}   &    00:16:59       &    07:54:10.1       &   +49:32:24.1        & 14.6     &  12\,475   \\
        \cline{5-9}               
                                 & &                        &                       &    00:17:29      &     07:54:19.3      &     +49:39:41.4       &  15.5    &  12\,112   \\
        \cline{3-9}
                                 & & \multirow{2}{*}{12}     & \multirow{2}{*}{567}   &    00:19:03       &    07:54:52.3       &   +50:06:10.6        & 18.9     &  11\,725   \\
        \cline{5-9}               
                                 & &                        &                       &    00:19:33      &     07:55:03.6      &     +50:15:25.6       &  20.3    &  11\,382   \\
        \cline{3-9}
                                 & & \multirow{2}{*}{13}     & \multirow{2}{*}{662}   &    00:20:08       &    07:55:18.9       &   +50:27:53.5        & 22.1     &  10\,954   \\
        \cline{5-9}               
                                 & &                        &                       &    00:20:38      &     07:55:32.5      &     +50:39:04.5       &  23.8    &  10\,600   \\
        \cline{3-9}
                                 & & \multirow{2}{*}{14}     & \multirow{2}{*}{915}   &    00:22:06       &    07:56:19.1       &   +51:17:58.0        & 30.4     &  9\,553   \\
        \cline{5-9}              
                                 & &                        &                       &    00:22:36      &     07:56:37.5        &    +51:33:25.7      &  33.2    &  9\,202   \\
        \hline
        \multirow{2}{*}{2024 EF} & \multirow{2}{*}{TRAPPIST-North}& \multirow{2}{*}{1}     & \multirow{2}{*}{485}   & 03:06:28     & 12:34:15.6       & +31:30:44.2        & 6.3     &  121\,308   \\
        \cline{5-9}
                                 & &                        &                      &    03:07:57      &  12:34:35.0       &  +31:22:14.8        & 6.4     &  120\,671   \\
        
    \hline \multicolumn{9}{l}{The length in pixel is the length of the trail in the image. For the Time, RA, DEC, Speed, and Range, the value at the beginning }\\
    \multicolumn{9}{l}{of the acquisition and at the end of the acquisition are provided for each frame. }\\
    \end{tabular}
    \label{tab:Observations}
\end{table*}

\section{Conclusions}

We presented in this paper a novel approach to perform photometry of fast moving near-Earth asteroids. Instead of tracking the asteroid motion, leading to stars appearing trailed on the images, we observed the asteroid using sidereal tracking and let the asteroid sweep through the field. This results in the asteroid appearing trailed on the images. 

The advantage of this technique is that it allows to extract the brightness of the object over time in single images by extracting the photometry using square apertures at different location of the trail. We show that this technique works and is highly effective in detecting fast rotating asteroids for which usual observation techniques would have failed to retrieve the period as their overhead time (read-out time of the CCD) is typically longer than the rotation period of the asteroid. 

We analyzed three recently observed targets, 2023 CX1, 2024 BX1, and 2024 EF. The first two impacted the Earth a few minutes after our observations ended while the third one performed a close fly-by.

For 2023 CX1, the photometric variation is less than 0.1 mag, suggesting the object was shaped nearly like a sphere, was seen pole-on, or did not rotate much during the span of our observations. A double-peaked phase curve is obtained for a spin period of 18.33 s. Such a short period is consistent with the small size of the asteroid derived from its absolute magnitude, but the significance of this result is low compared to the photometric measurement errors. 

In the case of 2024 BX1, we observed the fastest rotation period ever measured so far for an asteroid with a period of $P = 2.5888 \pm 0.0002$ s. In this case, the asteroid was elongated, resulting in a large amplitude of brightness variation.

In the case of 2024 EF, we observed a rotation period of approximately 3.95 min using regular asteroid tracked observations, but show that the photometry extracted on a single 90 s exposure trailed image can be correctly phased with the regular observations. This observation provides more detail to the shape of the lightcurve than the traditional method.

Based on these results, we encourage observers to obtain trailed observations of asteroids when the asteroids motion on the sky is so fast that the exposure time would be significantly shorter than the CCD read-out time. And especially when the asteroid is small and the expected rotation period is of the order of seconds to minutes. This is roughly equivalent to asteroids that are smaller than 10 m or H > 28. 

The use of trailed observations allowed us to obtain the fastest rotation period so far. However, asteroids are expected to rotate even faster. The limitation in detecting fast rotation periods arises from the use of exposure times larger than the rotation period. Fast spinning asteroids are expected to be small and thus faint when located far away from Earth. It is thus important to take advantage of their impacting and very close fly-by events to obtain reliable physical characterization on them.

\section*{Acknowledgements}

This research made use of Photutils, an Astropy package for detection and photometry of astronomical sources \citet{Photutils}.

TRAPPIST-North is a project funded by the University of Li\`{e}ge, in collaboration with the Cadi Ayyad University of Marrakech (Morocco) and the the Belgian Fund for Scientific Research (FNRS) under the grant PDR T.0120.21. E. Jehin is a FNRS Senior Research Associate.

This work made use of the NASA JPL Horizons ephemerides service available at \url{https://ssd.jpl.nasa.gov/horizons/}

PJ is supported by NASA's YORPD program (grant NNH21ZDA001N).



\bibliographystyle{aa}
\bibliography{adssample.bib}

\end{document}